\documentclass[prd,aps,nofootinbib,floatfix,10pt]{revtex4}

\usepackage{amsmath,graphicx,epsfig,amssymb,dsfont,mathtools,}
\usepackage[usenames]{color}
\usepackage{ulem} 
\usepackage{bigstrut}
\usepackage{slashed}
\usepackage{multirow}
\usepackage{makecell}
\usepackage{verbatim}
\usepackage{amsmath,graphicx,color,epsfig}
\usepackage{longtable}
  \allowdisplaybreaks[1]

\allowdisplaybreaks
\renewcommand{\thesubsection}{\Roman{subsection}}

\begin{document}
\title{Weak Decays of Doubly-Heavy Tetraquarks ${b\bar c}{q\bar q}$ }
\author{Gang Li~$^1$, Xiao-Feng Wang~$^1$, and
Ye Xing$^2$~\footnote{Email:xingye\_guang@sjtu.edu.cn}
}

\affiliation{
$^1$ School of Physics and Engineering, Qufu Normal University, Qufu 273165, China }
\affiliation{
 $^2$ INPAC,  SKLPPC, MOE Key Laboratory for Particle Physics, School of Physics and Astronomy, \\ Shanghai Jiao Tong University, Shanghai  200240, China}

\begin{abstract}
We study the weak decays of  exotic tetraquark states ${b\bar c}{q\bar q}$ with two heavy   quarks. Under the SU(3)  symmetry for light quarks,  these tetraquarks  can be classified into an octet plus a singlet: $3\bigotimes\bar 3=1\bigoplus8$. We will  concentrate on the octet tetraquarks  with  $J^{P}=0^{+}$, and study their weak decays,  both semileptonic and nonleptonic.  Hadron-level effective Hamiltonian is constructed according to the irreducible representations of the SU(3) group.  Expanding the Hamiltonian, we obtain the decay amplitudes parameterized in terms of a few  irreducible quantities. Based on these amplitudes, relations for decay widths are derived, which  can be tested in future. We also give a list of   golden channels that can be used to look for these states   at various colliders.
\end{abstract}

\maketitle

\section{Introduction}
Since the first discovery  of  X(3872) by Belle in 2003~\cite{Choi:2003ue}, a large number of charmonium-like and bottomonium-like hadrons have been discovered in the past decade~\cite{Tanabashi:2018oca}. Many
of these discovered  states defy a standard quarkonium interpretation and likely have a pair of  hidden flavored quarks, with the quark content $Q\bar Q q\bar q'$~(for a recent review, see Refs.~\cite{Chen:2016qju,Guo:2017jvc}). Here $Q$ represents a heavy bottom/charm quark and $q(q')$ denotes a light $u,d,s$ quark. Extensive theoretical studies have been carried out to explore their structures, productions and decays~\cite{Guo:2013sya,Cleven:2013sq,Guo:2013ufa,Liu:2016xly,Li:2014pfa,Li:2013xia,Guo:2014sca,Guo:2014ppa,Guo:2014iya,Chen:2016mjn,Wang:2013kra,Li:2012as,Li:2014uia,Albaladejo:2017blx,Liu:2013vfa,Guo:2013zbw,Voloshin:2013ez,Voloshin:2011qa,Chen:2011pv,Li:2013yla,Chen:2011pu,Chen:2012yr,Bondar:2011ev,Li:2015uwa,Chen:2013bha,Wu:2016ypc}.  In 2016, the D0 collaboration has reported an  evidence for the open-bottom tetraquark X(5568)~\cite{D0:2016mwd}, though it has not been confirmed by the other experimental groups~\cite{Aaij:2016iev,Sirunyan:2017ofq,Aaltonen:2017voc,Aaboud:2018hgx}. Therefore the existence of  open-flavored tetraquarks is an interesting question  in hadronic physics, in particular  the hadron spectroscopy.

Four-quark states with two different heavy quarks and two light quarks are of great interest  since they can provide a unique platform to study strong interactions under two color static sources. In the diquark-antidiquark model~\cite{Jaffe:2003sg}, the four-quarks system $[b q] [\bar c\bar q]$ with  orbital angular momentum L=0 can have $J^{P}=0^{+}$~\cite{Jaffe:2004ph}.  Since the $0^+$ tetraquarks are lowest lying, their weak decays  can provide unique insights to unravel their internal structure. In this paper, we adopt the SU(3) flavor symmetry   to handle   these weak decays. The SU(3) approach has been successfully applied into the B meson and heavy baryon decays~\cite{Savage:1989ub,Gronau:1995hm,He:1998rq,He:2000ys,Chiang:2004nm,Li:2007bh,Wang:2009azc,Cheng:2011qh,Hsiao:2015iiu,
Lu:2016ogy,He:2016xvd,Wang:2017vnc,Wang:2017mqp,Wang:2017azm,Shi:2017dto,Wang:2018utj,He:2018php} and a global  picture consistent data has been established.  In the SU(3) symmetry, the tetraquarks with two light quarks can form an octet and a singlet. In this work, we will concentrate on the octet,  abbreviated as $X_{b\bar c8}$.

In the following, we will first construct the hadron-level effective Hamiltonian   according to  the irreducible representations of the SU(3) group.  Expanding these  Hamiltonian, we obtain the decay amplitudes parameterized in terms of a few SU(3) irreducible quantities. Based on the expanded amplitudes, relations for decay widths are derived, which  can be examined  in future. We also give a list of   golden channels that can be used to look for these states   at various colliders.

The rest of this paper is organized as follows. In Sec. II, we give the multiplets expressions under the SU(3) flavor symmetry. From Sec. III to Sec. IV, we mainly study the semi-leptonic and non-leptonic weak decays of the $X_{b\bar c8}$ states. In Sec. V, we will give a collection of the golden channels that can be used to discovery the doubly heavy tetraquarks  in future experiments. we summarize in the last section.

\section{Particle Multiplets}
\label{sec:particle_multiplet}

Based on the light flavor SU(3) symmetry, open-flavor tetraquark with the quark constituents ${b\bar c}{q\bar q}$ can form an octet and a singlet, of which the octet can be represented as
\begin{eqnarray}
 T_{\bf{b\bar{c} 8}}= \left(\begin{array}{ccc}   \frac{T^{Bc}_{\pi^0}}{\sqrt{2}}+\frac{T^{Bc}_{\eta}}{\sqrt{6}} & T^{Bc}_{\pi^+} &  T^{Bc}_{K^+} \\ T^{Bc}_{\pi^-}&  -\frac{T^{Bc}_{\pi^0}}{\sqrt{2}}+\frac{T^{Bc}_{\eta}}{\sqrt{6}} & T^{Bc}_{K^0} \\ T^{Bc}_{K^-}  &  T^{Bc}_{\overline K^0}  & -\frac{2}{\sqrt{6}}T^{Bc}_{\eta}
  \end{array} \right)\,.
\end{eqnarray}
For simplicity, we will  not consider the flavor singlet in this work. The decomposition can be reached by $3\bigotimes \bar 3=1\bigoplus 8$.

In the meson sector, light pseudoscalar mesons or vector mesons can also form an octet plus a singlet, generally, the octet is written as
\begin{eqnarray}
 M_{8}=\begin{pmatrix}
 \frac{\pi^0}{\sqrt{2}}+\frac{\eta}{\sqrt{6}}
 &\pi^+ & K^+\\
 \pi^-&-\frac{\pi^0}{\sqrt{2}}+\frac{\eta}{\sqrt{6}}&{K^0}\\
 K^-&\bar K^0 &-2\frac{\eta}{\sqrt{6}}
 \end{pmatrix},
\end{eqnarray}
as the same quark content, vector meson octet will give the similar structure. Besides, we need the representation of bottom mesons which form an SU(3) anti-triplet given as: $B_i=\left(\begin{array}{ccc} B^-, & \overline B^0, &\overline B^0_s  \end{array} \right),$
and the representation of anti-triplet charmed mesons given as: $D_i=\left(\begin{array}{ccc} D^0, & D^+, & D^+_s  \end{array} \right), \;\;\;
\overline D^i=\left(\begin{array}{ccc}\overline D^0, & D^-, & D^-_s  \end{array} \right).$


\section{Semi-Leptonic $T_{b\bar c8}$ decays}
\renewcommand\thesubsection{(\Roman{subsection})}
\label{sec:semileptonic_decay}

In this section, we will discuss the possible semi-leptonic weak decay modes of the octet tetraquark $T_{b\bar c8}$. Considering the decay modes at quark level, $T_{b\bar c8}$ will hold both $b$-quark and $\bar c$-quark decays. For the $b$-quark,  semi-leptonic weak decays are governed by
\begin{eqnarray}
b\to c/u \ell^- \bar \nu_{\ell}.
\end{eqnarray}
For the $\bar c$-quark, semi-leptonic decays are induced by
\begin{eqnarray}
  \bar c\to  \bar d/\bar s  \ell^-   \bar \nu_{\ell}.
\end{eqnarray}
In the following, we will study the decays above in order.
\begin{figure}
\includegraphics[width=0.5\columnwidth]{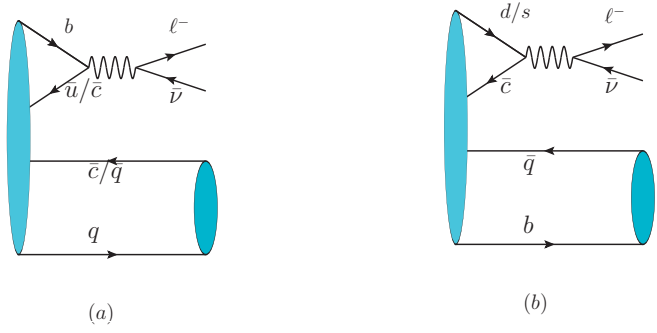}
\caption{Feynman diagrams for semileptonic decays of tetraquark $X_{b\bar c 8}$. Panel (a) corresponds to the $b$ quark decay and panel (b) denotes the $\bar c$ quark decay. }
\label{fig:topology-1}
\end{figure}

\subsubsection{$b\to c/u \ell^- \overline \nu_{\ell}$:  decays into a meson and $\ell^- \overline \nu_{\ell}$}
In $b$ quark decay, the general electro-weak  Hamiltonian  of the $b\to c/u \ell^- \overline \nu_{\ell}$ transition can be expressed as
\begin{eqnarray}
 {\cal H}_{eff} &=& \frac{G_F}{\sqrt2} \left[V_{q'b} \bar q' \gamma^\mu(1-\gamma_5)b \bar  \ell\gamma_\mu(1-\gamma_5) \nu_{\ell}\right] +h.c.,
\end{eqnarray}
with $q'=u,c$, in which the electro-weak vertex is suggested to be a $V-A$ structure. As a contrast, the vertex forms a triplet representation $H_{3}'$  within SU(3) flavor symmetry, specifically $(H_3')^1=1$ and $(H_3')^{2,3}=0$. At the hadron level, the transition can be included into the process that $X_{b\bar c 8}$ decays to a charmed meson and $\ell \overline \nu_{\ell}$.
Following the SU(3) analysis, the Hamiltonian of hadronic level is constructed as
\begin{eqnarray}
&\mathcal{H}_{eff}&=a_1(T_{b\bar c8})^i_j (H_3')^j D_i \bar \ell \nu\;,
\end{eqnarray}
here, the coefficient $a_{1}$ represents the non-perturbative parameter. For completeness, we give the corresponding Feynman diagram at quark level shown in Fig.~\ref{fig:topology-1}.(a). It is convenient to achieve the decay amplitudes given in Tab.~\ref{tab:bc8_meson_lv} by expanding the Hamiltonian constructed above, in which all amplitudes are represented as $a_1$. Therefor, we can directly obtain the relations between different decay channels given as follows.
\begin{eqnarray*}
\Gamma(X^{Bc}_{\pi^-}\to D^- l^-\bar\nu)=2\Gamma(X^{Bc}_{\pi^0}\to \overline D^0 l^-\bar\nu )=\Gamma(X^{Bc}_{K^-}\to  D^-_s l^-\bar\nu )=6\Gamma(X^{Bc}_{\eta}\to \overline D^0 l^-\bar\nu).
\end{eqnarray*}
It is should be note that the phase space difference will provide corrections to these relations.

\begin{table}
\caption{Amplitudes for tetraquark $X_{b \bar{c} 8}$ decays into anti-charmed meson or a light meson.}\label{tab:bc8_meson_lv}\begin{tabular}{|cc|cc|}\hline\hline
channel & amplitude&channel & amplitude \\\hline
$X^{Bc}_{\pi^-}\to D^- l^-\bar\nu $ & $ a_1 V_{\text{ub}}$&
$X^{Bc}_{\pi^0}\to \overline D^0 l^-\bar\nu $ & $ \frac{a_1 V_{\text{ub}}}{\sqrt{2}}$\\\hline
$X^{Bc}_{K^-}\to  D^-_s l^-\bar\nu $ & $ a_1 V_{\text{ub}}$&
$X^{Bc}_{\eta}\to \overline D^0 l^-\bar\nu $ & $ \frac{a_1 V_{\text{ub}}}{\sqrt{6}}$\\\hline
\hline
channel & amplitude &channel & amplitude \\\hline
$X^{Bc}_{\pi^-}\to \pi^-  l^-\bar\nu $ & $ a_2$&
$X^{Bc}_{\pi^0}\to \pi^0  l^-\bar\nu $ & $ a_2$\\\hline
$X^{Bc}_{\pi^+}\to \pi^+  l^-\bar\nu $ & $ a_2$&
$X^{Bc}_{K^-}\to K^-  l^-\bar\nu $ & $ a_2$\\\hline
$X^{Bc}_{\overline K^0}\to \overline K^0  l^-\bar\nu $ & $ a_2$&
$X^{Bc}_{K^0}\to K^0  l^-\bar\nu $ & $ a_2$\\\hline
$X^{Bc}_{K^+}\to K^+  l^-\bar\nu $ & $ a_2$&
$X^{Bc}_{\eta}\to \eta  l^-\bar\nu $ & $ a_2$\\\hline
\hline
\end{tabular}
\end{table}
For the SU(3) singlet $b\to c$ transition, the process at the hadron level is that $X_{b\bar c8}$ decays into a light meson octet and $\ell \overline \nu_{\ell} $. Consequently, the Hamiltonian at the hadron level is constructed as
\begin{eqnarray}
&\mathcal{H}_{eff}&=a_2 (T_{b\bar c8})^i_j M^j_i \bar \ell \nu \;.
\end{eqnarray}
One then obtain the amplitudes of different decay channels listed in Tab.~\ref{tab:bc8_meson_lv}, from which we derive that all the channels in the transition give the equal decay widths.

\subsubsection{$\bar c\to \bar d/\bar s \ell^-  \bar \nu$  decay into B meson and $\ell^-  \bar \nu$}

In $\bar c$ quark decay, the electro-weak effective Hamiltonian are given as
\begin{eqnarray}
{\cal H}_{eff}&=&\frac{G_F}{\sqrt2} \left[V_{cq} \bar c  \gamma^\mu(1-\gamma_5)q \bar \ell \gamma_\mu(1-\gamma_5) \nu_{\ell}\right] +h.c.,
\end{eqnarray}
where $q=d,s$, $V_{cd}$ and $V_{cs}$  are CKM matrix elements. Under the SU(3) symmetry, the $\bar c\to \bar q \ell^-\bar \nu$ transition can form an SU(3) triplet vertex, denoted as $H_{  3}$ with $(H_{  3})^1=0,~(H_{  3})^2=V_{cd},~(H_{  3})^3=V_{cs}$.
Consistently, We construct the Hamiltonian at the hadron level as follows.
\begin{eqnarray}
&\mathcal{H}_{eff}&=c_1(T_{b\bar c8})^i_j (H_3)_i \overline B^j  \bar \ell \nu\;.
\end{eqnarray}
The decay amplitudes deduced from the Hamiltonian above are listed in Tab.~\ref{tab:bc8_B_lv}. For completeness, we give the corresponding Feynman diagrams given in Fig.~\ref{fig:topology-1}.(b). One then obtain the relations between different channels as follows.
\begin{eqnarray*}
&&\Gamma(X^{Bc}_{\pi^-}\to B^- l^-\bar\nu )=2\Gamma(X^{Bc}_{\pi^0}\to \overline B^0 l^-\bar\nu)=\Gamma(X^{Bc}_{K^0}\to \overline B^0_s l^-\bar\nu)=6\Gamma(X^{Bc}_{\eta}\to \overline B^0 l^-\bar\nu),\\
&&\Gamma(X^{Bc}_{K^-}\to B^- l^-\bar\nu )=\Gamma(X^{Bc}_{\overline K^0}\to \overline B^0 l^-\bar\nu)=\frac{3}{2}\Gamma(X^{Bc}_{\eta}\to \overline B^0_s l^-\bar\nu).
\end{eqnarray*}
More technically, the relations between different channels may be modified with a view to the SU(3) symmetry breaking effects in the charmed or anti-charmed quark decays.
\begin{table}
\caption{Amplitudes for tetraquark $X_{b\bar c8}$  decays into a B meson.}\label{tab:bc8_B_lv}
\begin{tabular}{|cc|cc|}\hline\hline
channel & amplitude&channel & amplitude \\\hline
$X^{Bc}_{\pi^-}\to B^- l^-\bar\nu $ & $ c_1 V_{\text{cd}}{}$&
$X^{Bc}_{\pi^0}\to \overline B^0 l^-\bar\nu $ & $ -\frac{c_1 V_{\text{cd}}{}}{\sqrt{2}}$\\\hline
$X^{Bc}_{K^-}\to B^- l^-\bar\nu $ & $ c_1 V_{\text{cs}}{}$&
$X^{Bc}_{\overline K^0}\to \overline B^0 l^-\bar\nu $ & $ c_1 V_{\text{cs}}{}$\\\hline
$X^{Bc}_{K^0}\to \overline B^0_s l^-\bar\nu $ & $ c_1 V_{\text{cd}}{}$&
$X^{Bc}_{\eta}\to \overline B^0 l^-\bar\nu $ & $ \frac{c_1 V_{\text{cd}}{}}{\sqrt{6}}$\\\hline
$X^{Bc}_{\eta}\to \overline B^0_s l^-\bar\nu $ & $ -\sqrt{\frac{2}{3}} c_1 V_{\text{cs}}{}$& &\\
\hline\hline
\end{tabular}
\end{table}

\section{Non-Leptonic $T_{b\bar c8}$ decays}
\label{sec:nonleptonic_decay}

\begin{figure}
\includegraphics[width=0.8\columnwidth]{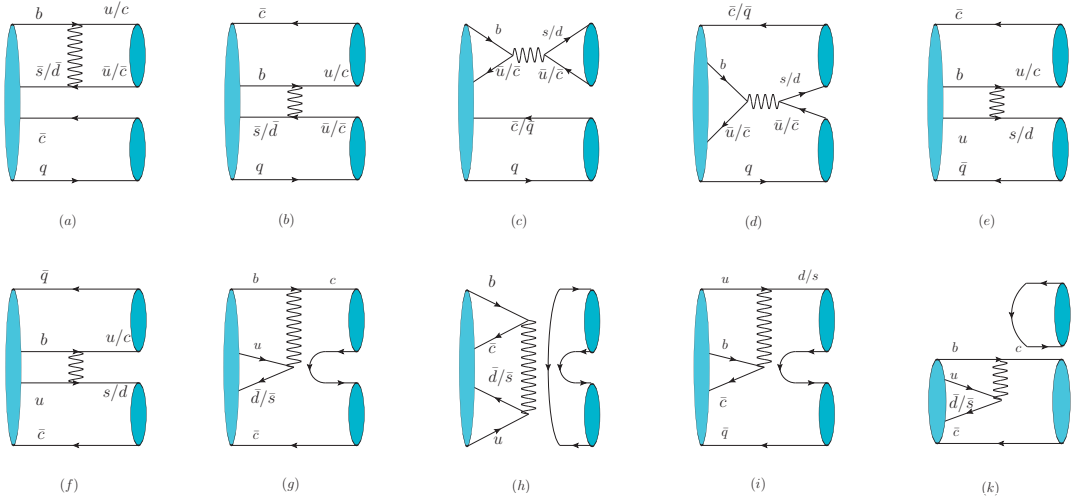}
\caption{Feynman diagrams for the $b$-quark non-leptonic decays of tetraquark $X_{b\bar c8}$. Panels (a-k) correspond to the decays into a pair of mesons. In panels (k), the final meson produced by gluons is the flavor singly state which we will not consider here. The diagrams in panels(c,d,g,h,i) are usually power suppressed as a pair of quark and anti-quark in the initial state can annihilate. }
\label{fig:topology-2}
\end{figure}
\begin{figure}
\includegraphics[width=0.8\columnwidth]{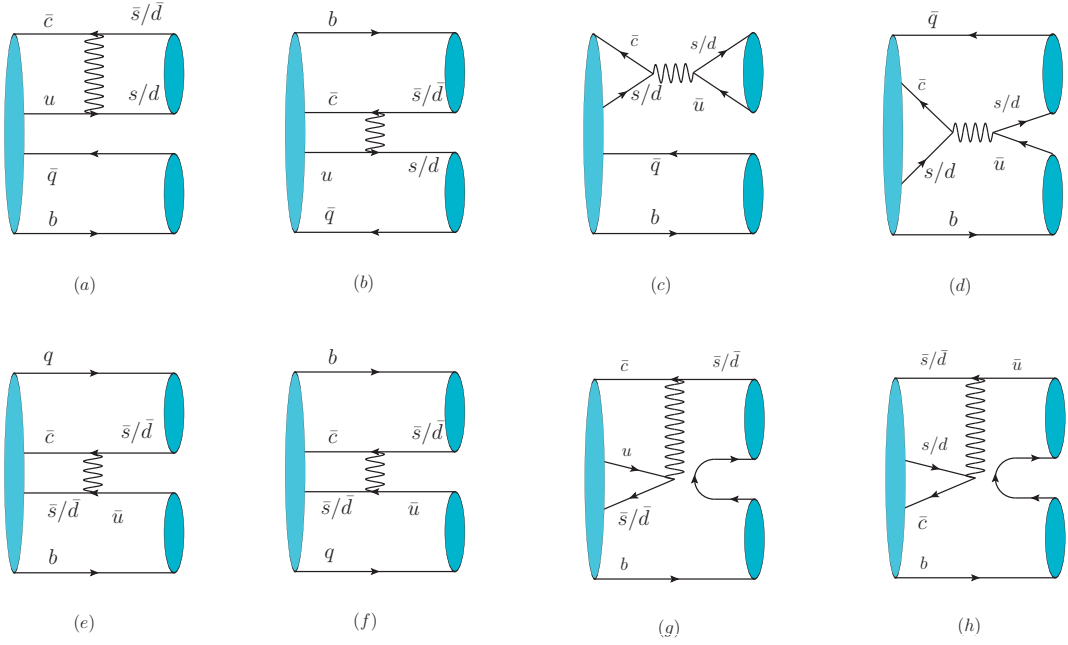}
\caption{Feynman diagrams for the $\bar c$-quark non-leptonic decays of tetraquark $X_{b\bar c8}$. The two-body processes are given in panels(a-h). The $\bar c$-quark decays have the similar structures with the $b$-quark decays. The panels(a-h) contribute to the process of $X_{b\bar c8}$ decays into $B$ plus a light meson.  }
\label{fig:topology-4}
\end{figure}

In $b$ quark non-leptonic decays, the transitions can be classified into four different kinds in the light of CKM matrix:
\begin{eqnarray}
b\to c\bar c d/s, \;
b\to c \bar u d/s, \;
b\to u \bar c d/s, \;
b\to q_1 \bar q_2 q_3,
\end{eqnarray}
here $q_{1,2,3}$ represent the light quark.

In $\bar c$ quark non-leptonic decays, the pronounced classifications are given as:
\begin{eqnarray}
\bar c\to \bar s d \bar u, \;
\bar c\to \bar u d \bar d/s \bar s, \;
\bar c\to \bar d s \bar u, \;
\end{eqnarray}
which are Cabibbo allowed, singly Cabibbo suppressed, and doubly Cabibbo suppressed respectively.
In the following, we will study the $X_{b\bar c8}$ non-leptonic decays in these orders.

\subsection{$b\to c\bar c d/s$ transition: two-body decays into mesons}

The operator of the $b\to c\bar c d/s$  transition can form an triplet under the SU(3) light quark symmetry, according to that we can write down the effective Hamiltonian of  $X_{b\bar c8}$ producing two mesons as follows.
\begin{eqnarray}
&\mathcal{H}_{eff}&=a_1 (T_{b\bar c8})^i_j (H_3)^j D_i J/\psi,\nonumber\\
&\mathcal{H}_{eff}&=a_2 (T_{b\bar c8})^i_j (H_3)^j D_k M^k_i + a_3 (T_{b\bar c8})^i_j (H_3)^k D_k M^j_i +a_4 (T_{b\bar c8})^i_j (H_3)^k D_i M^j_k,
\end{eqnarray}
with $(H_{  3})^{2}=V_{cd}^*$ and $(H_{  3})^{3}=V_{cs}^*$. Consistently, the corresponding Feynman diagrams are given in  Fig.~\ref{fig:topology-2}(a-d). In particular, the diagrams in  Fig.~\ref{fig:topology-2}(a,b) represent $X_{b\bar c8}$ decays into $D$ and $J/\psi$ mesons, and the diagrams in  Fig.~\ref{fig:topology-2}(c,d) denote processes with $D$ and light mesons final states.  Expanding the two Hamiltonian above, one obtains the decay amplitudes which are listed in Tab.~\ref{tab:bc8_Dbar_Jpsi_Dbar_M}. In addition, the relations between the different decay widths are given as follows.
\begin{eqnarray*}
\Gamma(X^{Bc}_{\pi^0}\to K^- \overline D^0)= \frac{1}{2}\Gamma(X^{Bc}_{\pi^-}\to K^- D^-)=\Gamma(X^{Bc}_{\pi^0}\to \overline K^0 D^-)=\frac{1}{2}\Gamma(X^{Bc}_{\pi^+}\to \overline K^0 \overline D^0),\\
\Gamma(X^{Bc}_{\pi^+}\to \eta \overline D^0)= 2\Gamma(X^{Bc}_{\pi^0}\to \eta D^-)=
 2\Gamma(X^{Bc}_{\eta}\to \pi^0 D^-)=\Gamma(X^{Bc}_{\eta}\to \pi^- \overline D^0),\\
 \Gamma(X^{Bc}_{K^0}\to \pi^- \overline D^0)= 2\Gamma(X^{Bc}_{K^0}\to \pi^0 D^-)=
\Gamma(X^{Bc}_{K^+}\to \pi^+ D^-)=2\Gamma(X^{Bc}_{K^+}\to \pi^0 \overline D^0),\\
\Gamma(X^{Bc}_{\overline K^0}\to K^- \overline D^0)= 2\Gamma(X^{Bc}_{\pi^0}\to K^0  D^-_s)=
\Gamma(X^{Bc}_{\pi^+}\to K^+  D^-_s),\\
\Gamma(X^{Bc}_{\pi^0}\to \pi^0  D^-_s)= { }\Gamma(X^{Bc}_{\pi^-}\to \pi^-  D^-_s)=\Gamma(X^{Bc}_{\pi^+}\to \pi^+  D^-_s),\\
\Gamma(X^{Bc}_{K^+}\to K^0 \overline D^0)= { }\Gamma(X^{Bc}_{K^-}\to \pi^-  D^-_s)=
2\Gamma(X^{Bc}_{\overline K^0}\to \pi^0  D^-_s),\\
\Gamma(X^{Bc}_{K^+}\to \eta \overline D^0)= { }\Gamma(X^{Bc}_{K^0}\to \eta D^-),
\Gamma(X^{Bc}_{\eta}\to K^- \overline D^0)= { }\Gamma(X^{Bc}_{\eta}\to \overline K^0 D^-),\\
\Gamma(X^{Bc}_{\pi^-}\to \pi^- D^-)= { }\Gamma(X^{Bc}_{K^0}\to K^0 D^-),
\Gamma(X^{Bc}_{\pi^+}\to \pi^+ D^-)= { }\Gamma(X^{Bc}_{\overline K^0}\to \overline K^0 D^-),\\
\Gamma(X^{Bc}_{\pi^+}\to \pi^0 \overline D^0)= { }\Gamma(X^{Bc}_{\pi^0}\to \pi^- \overline D^0),
\Gamma(X^{Bc}_{\overline K^0}\to \overline K^0  D^-_s)= { }\Gamma(X^{Bc}_{K^-}\to K^-  D^-_s),\\
\Gamma(X^{Bc}_{K^+}\to K^+ D^-)= { }\Gamma(X^{Bc}_{K^-}\to K^- D^-),
\Gamma(X^{Bc}_{K^+}\to K^+  D^-_s)= { }\Gamma(X^{Bc}_{K^0}\to K^0  D^-_s),\\
2\Gamma(X^{Bc}_{\pi^0}\to   D^-  J/\psi)=\Gamma(X^{Bc}_{\pi^+}\to   \overline D^0  J/\psi )=\Gamma(X^{Bc}_{\overline K^0}\to    D^-_s  J/\psi )=6\Gamma(X^{Bc}_{\eta}\to   D^-  J/\psi),\\
\Gamma(X^{Bc}_{K^0}\to   D^-  J/\psi)=\Gamma(X^{Bc}_{K^+}\to   \overline D^0  J/\psi )=\frac{3}{2}\Gamma(X^{Bc}_{\eta}\to    D^-_s  J/\psi).
\end{eqnarray*}
\begin{table}
\caption{Tetraquark $X_{b\bar c8}$ decays into anti-charmed meson plus light meson or anti-charmed meson plus $J/\psi$.}\label{tab:bc8_Dbar_Jpsi_Dbar_M}\begin{tabular}{|cc|cc|}\hline\hline
channel & amplitude&channel & amplitude \\\hline
$X^{Bc}_{\pi^-}\to   \pi^-   D^- $ & $ \left(a_3+a_4\right) V_{cd}^*$&
$X^{Bc}_{\pi^-}\to   \pi^-    D^-_s $ & $ a_3 V_{cs}^*$\\\hline
$X^{Bc}_{\pi^-}\to   K^-   D^- $ & $ a_4 V_{cs}^*$&
$X^{Bc}_{\pi^0}\to   \pi^0   D^- $ & $ \frac{1}{2} \left(a_2+2 a_3+a_4\right) V_{cd}^*$\\\hline
$X^{Bc}_{\pi^0}\to   \pi^0    D^-_s $ & $ a_3 V_{cs}^*$&
$X^{Bc}_{\pi^0}\to   \pi^-   \overline D^0 $ & $ \frac{\left(a_4-a_2\right) V_{cd}^*}{\sqrt{2}}$\\\hline
$X^{Bc}_{\pi^0}\to   K^0    D^-_s $ & $ -\frac{a_2 V_{cd}^*}{\sqrt{2}}$&
$X^{Bc}_{\pi^0}\to   \overline K^0   D^- $ & $ -\frac{a_4 V_{cs}^*}{\sqrt{2}}$\\\hline
$X^{Bc}_{\pi^0}\to   K^-   \overline D^0 $ & $ \frac{a_4 V_{cs}^*}{\sqrt{2}}$&
$X^{Bc}_{\pi^0}\to   \eta   D^- $ & $ -\frac{\left(a_2+a_4\right) V_{cd}^*}{2 \sqrt{3}}$\\\hline
$X^{Bc}_{\pi^+}\to   \pi^+   D^- $ & $ \left(a_2+a_3\right) V_{cd}^*$&
$X^{Bc}_{\pi^+}\to   \pi^+    D^-_s $ & $ a_3 V_{cs}^*$\\\hline
$X^{Bc}_{\pi^+}\to   \pi^0   \overline D^0 $ & $ \frac{\left(a_2-a_4\right) V_{cd}^*}{\sqrt{2}}$&
$X^{Bc}_{\pi^+}\to   K^+    D^-_s $ & $ a_2 V_{cd}^*$\\\hline
$X^{Bc}_{\pi^+}\to   \overline K^0   \overline D^0 $ & $ a_4 V_{cs}^*$&
$X^{Bc}_{\pi^+}\to   \eta   \overline D^0 $ & $ \frac{(a_2 +a_4) V_{cd}^*}{\sqrt{6}}$\\\hline
$X^{Bc}_{K^-}\to   \pi^-    D^-_s $ & $ a_4 V_{cd}^*$&
$X^{Bc}_{K^-}\to   K^-   D^- $ & $ a_3 V_{cd}^*$\\\hline
$X^{Bc}_{K^-}\to   K^-    D^-_s $ & $ \left(a_3+a_4\right) V_{cs}^*$&
$X^{Bc}_{\overline K^0}\to   \pi^0    D^-_s $ & $ -\frac{a_4 V_{cd}^*}{\sqrt{2}}$\\\hline
$X^{Bc}_{\overline K^0}\to   \overline K^0   D^- $ & $ \left(a_2+a_3\right) V_{cd}^*$&
$X^{Bc}_{\overline K^0}\to   \overline K^0    D^-_s $ & $ \left(a_3+a_4\right) V_{cs}^*$\\\hline
$X^{Bc}_{\overline K^0}\to   K^-   \overline D^0 $ & $ a_2 V_{cd}^*$&
$X^{Bc}_{\overline K^0}\to   \eta    D^-_s $ & $ \frac{\left(a_4-2 a_2\right) V_{cd}^*}{\sqrt{6}}$\\\hline
$X^{Bc}_{K^0}\to   \pi^0   D^- $ & $ -\frac{a_2 V_{cs}^*}{\sqrt{2}}$&
$X^{Bc}_{K^0}\to   \pi^-   \overline D^0 $ & $ a_2 V_{cs}^*$\\\hline
$X^{Bc}_{K^0}\to   K^0   D^- $ & $ \left(a_3+a_4\right) V_{cd}^*$&
$X^{Bc}_{K^0}\to   K^0    D^-_s $ & $ \left(a_2+a_3\right) V_{cs}^*$\\\hline
$X^{Bc}_{K^0}\to   \eta   D^- $ & $ \frac{\left(a_2-2 a_4\right) V_{cs}^*}{\sqrt{6}}$&
$X^{Bc}_{K^+}\to   \pi^+   D^- $ & $ a_2 V_{cs}^*$\\\hline
$X^{Bc}_{K^+}\to   \pi^0   \overline D^0 $ & $ \frac{a_2 V_{cs}^*}{\sqrt{2}}$&
$X^{Bc}_{K^+}\to   K^+   D^- $ & $ a_3 V_{cd}^*$\\\hline
$X^{Bc}_{K^+}\to   K^+    D^-_s $ & $ \left(a_2+a_3\right) V_{cs}^*$&
$X^{Bc}_{K^+}\to   K^0   \overline D^0 $ & $ a_4 V_{cd}^*$\\\hline
$X^{Bc}_{K^+}\to   \eta   \overline D^0 $ & $ \frac{\left(a_2-2 a_4\right) V_{cs}^*}{\sqrt{6}}$&
$X^{Bc}_{\eta}\to   \pi^0   D^- $ & $ -\frac{\left(a_2+a_4\right) V_{cd}^*}{2 \sqrt{3}}$\\\hline
$X^{Bc}_{\eta}\to   \pi^-   \overline D^0 $ & $ \frac{(a_2 +a_4) V_{cd}^*}{\sqrt{6}}$&
$X^{Bc}_{\eta}\to   K^0    D^-_s $ & $ \frac{\left(a_2-2 a_4\right) V_{cd}^*}{\sqrt{6}}$\\\hline
$X^{Bc}_{\eta}\to   \overline K^0   D^- $ & $ \frac{\left(a_4-2 a_2\right) V_{cs}^*}{\sqrt{6}}$&
$X^{Bc}_{\eta}\to   K^-   \overline D^0 $ & $ \frac{\left(a_4-2 a_2\right) V_{cs}^*}{\sqrt{6}}$\\\hline
$X^{Bc}_{\eta}\to   \eta   D^- $ & $ \frac{1}{6} \left(a_2+6 a_3+a_4\right) V_{cd}^*$&
$X^{Bc}_{\eta}\to   \eta    D^-_s $ & $ \frac{1}{3} \left(2 a_2+3 a_3+2 a_4\right) V_{cs}^*$\\\hline
\hline
channel & amplitude&channel & amplitude \\\hline
$X^{Bc}_{\pi^0}\to   D^-  J/\psi $ & $ -\frac{a_1 V_{cd}^*}{\sqrt{2}}$&
$X^{Bc}_{\pi^+}\to   \overline D^0  J/\psi $ & $ a_1 V_{cd}^*$\\\hline
$X^{Bc}_{\overline K^0}\to    D^-_s  J/\psi $ & $ a_1 V_{cd}^*$&
$X^{Bc}_{K^0}\to   D^-  J/\psi $ & $ a_1 V_{cs}^*$\\\hline
$X^{Bc}_{K^+}\to   \overline D^0  J/\psi $ & $ a_1 V_{cs}^*$&
$X^{Bc}_{\eta}\to   D^-  J/\psi $ & $ \frac{a_1 V_{cd}^*}{\sqrt{6}}$\\\hline
$X^{Bc}_{\eta}\to    D^-_s  J/\psi $ & $ -\sqrt{\frac{2}{3}} a_1 V_{cs}^*$& &\\
\hline
\end{tabular}
\end{table}

\subsection{$b\to c \bar u d/s$ transition: two body decays into mesons}


The operator of the $b\to c \bar u d/s$ transition can form an octet according to the SU(3) symmetry, of which nonzero entry
$(H_{{\bf8}})^2_1 =V_{ud}^*$ for   the $b\to c\bar ud$  transition, and   $(H_{{\bf8}})^3_1 =V_{us}^*$ for  the $b\to c\bar
us$ transition. As usual, the hadron-level effective Hamiltonian can be constructed as
\begin{eqnarray}
&\mathcal{H}_{eff}&=a_6(T_{b\bar c8})^i_j (H_8)^k_i M^j_k J/\psi + a_7(T_{b\bar c8})^i_j (H_8)^j_k M^k_i J/\psi  ,\nonumber\\
&\mathcal{H}_{eff}&=a_8(T_{b\bar c8})^i_j (H_8)^j_i D_k \overline D^k +a_9 (T_{b\bar c8})^i_j (H_8)^k_i D_k \overline D^j +a_{10}(T_{b\bar c8})^i_j (H_8)^j_k D_i \overline D^k ,\nonumber\\
&\mathcal{H}_{eff}&=a_{11}(T_{b\bar c8})^i_j (H_8)^j_i M^l_k M^k_l + a_{12}(T_{b\bar c8})^i_j (H_8)^l_i M^j_k M^k_l + a_{13}(T_{b\bar c8})^i_j (H_8)^j_k M^l_i M^k_l \nonumber\\
&&+ a_{14}(T_{b\bar c8})^i_j (H_8)^l_k M^j_i M^k_l + a_{15}(T_{b\bar c8})^i_j (H_8)^l_k M^k_i M^j_l.
\end{eqnarray}
In topological level, the relevant Feynman diagrams are shown in Fig.~\ref{fig:topology-2}(a-i). Specifically, the Feynman diagrams of $X_{b\bar c8}$ decays into a light meson plus $J/\psi$ are given in Fig.~\ref{fig:topology-2}(b,e), and the Fig.~\ref{fig:topology-2}(a,f,g) correspond with the processes of producing $D$ plus $\bar D$, the processes that $X_{b\bar c8}$ produces two light mesons are represented by Feynman diagrams in Fig.~\ref{fig:topology-2}(c,d,h,i). One derives the decay amplitudes given in Tab.~\ref{tab:bc8_Jpsi_M_D_Dbar_M_M}  respectively. Accordingly, we obtain the relations between different decay widths for $J/\psi$ and a light meson as follows.
\begin{eqnarray*}
    \Gamma(X^{Bc}_{\pi^+}\to \pi^0 J/\psi)= { }\Gamma(X^{Bc}_{\pi^0}\to \pi^- J/\psi)=2\Gamma(X^{Bc}_{\pi^0}\to K^- J/\psi),\\
    \Gamma(X^{Bc}_{\pi^+}\to \overline K^0 J/\psi)= 2\Gamma(X^{Bc}_{\pi^0}\to K^- J/\psi),\\
    \Gamma(X^{Bc}_{K^+}\to \pi^0 J/\psi)= \frac{1}{2}\Gamma(X^{Bc}_{K^0}\to \pi^- J/\psi).
\end{eqnarray*}
The relations for producing the charmed meson and anti-charmed meson become
\begin{eqnarray*}
    \Gamma(X^{Bc}_{\pi^0}\to  D^0 D^-_s)= \frac{1}{2}\Gamma(X^{Bc}_{\pi^+}\to  D^+ D^-_s).
\end{eqnarray*}
The relations for producing two light mesons become
\begin{eqnarray*}
\Gamma(X^{Bc}_{\pi^-}\to \pi^- \pi^- )= 2\Gamma(X^{Bc}_{K^-}\to K^- \pi^- )= 2\Gamma(X^{Bc}_{K^0}\to K^0 \pi^- )=2\Gamma(X^{Bc}_{\pi^0}\to \pi^0 \pi^- ),\\
\Gamma(X^{Bc}_{\pi^-}\to \pi^- K^- )= \frac{1}{2}\Gamma(X^{Bc}_{K^-}\to K^- K^- )=
\Gamma(X^{Bc}_{\overline K^0}\to \overline K^0 K^- ),\\
\Gamma(X^{Bc}_{\pi^0}\to \pi^- \overline K^0 )= 3\Gamma(X^{Bc}_{\pi^+}\to \eta \overline K^0 )=
\Gamma(X^{Bc}_{\pi^+}\to \pi^0 \overline K^0 ),\\
\Gamma(X^{Bc}_{K^+}\to \pi^0 \pi^0 )= \frac{1}{2}\Gamma(X^{Bc}_{K^+}\to \pi^+ \pi^- ),
\Gamma(X^{Bc}_{\pi^+}\to \pi^0 \eta )= { }\Gamma(X^{Bc}_{\pi^0}\to \pi^- \eta ),\\
\Gamma(X^{Bc}_{\pi^+}\to \overline K^0 \eta )= 2\Gamma(X^{Bc}_{\pi^0}\to K^- \eta ),
\Gamma(X^{Bc}_{\overline K^0}\to \pi^0 K^- )= 3\Gamma(X^{Bc}_{\overline K^0}\to \eta K^- ),\\
\Gamma(X^{Bc}_{K^+}\to \pi^0 K^0 )= 3\Gamma(X^{Bc}_{K^+}\to \eta K^0 ),
\Gamma(X^{Bc}_{K^+}\to \pi^0 \eta )= \frac{1}{2}\Gamma(X^{Bc}_{K^0}\to \pi^- \eta ).
\end{eqnarray*}

\begin{table}
\caption{Tetraquark $X_{b\bar c8}$ decays into $J/\psi$ plus light meson or charmed meson plus anti-charmed meson or two light mesons.}\label{tab:bc8_Jpsi_M_D_Dbar_M_M}\begin{tabular}{|cc|cc|}\hline\hline
channel & amplitude &channel & amplitude\\\hline
$X^{Bc}_{\pi^0}\to   \pi^-   J/\psi $ & $ \frac{\left(a_6-a_7\right) V_{ud}^*}{\sqrt{2}}$&
$X^{Bc}_{\pi^0}\to   K^-   J/\psi $ & $ \frac{a_6 V_{us}^*}{\sqrt{2}}$\\\hline
$X^{Bc}_{\pi^+}\to   \pi^0   J/\psi $ & $ \frac{\left(a_7-a_6\right) V_{ud}^*}{\sqrt{2}}$&
$X^{Bc}_{\pi^+}\to   \overline K^0   J/\psi $ & $ a_6 V_{us}^*$\\\hline
$X^{Bc}_{\pi^+}\to   \eta   J/\psi $ & $ \frac{\left(a_6+a_7\right) V_{ud}^*}{\sqrt{6}}$&
$X^{Bc}_{\overline K^0}\to   K^-   J/\psi $ & $ a_7 V_{ud}^*$\\\hline
$X^{Bc}_{K^0}\to   \pi^-   J/\psi $ & $ a_7 V_{us}^*$&
$X^{Bc}_{K^+}\to   \pi^0   J/\psi $ & $ \frac{a_7 V_{us}^*}{\sqrt{2}}$\\\hline
$X^{Bc}_{K^+}\to   K^0   J/\psi $ & $ a_6 V_{ud}^*$&
$X^{Bc}_{K^+}\to   \eta   J/\psi $ & $ \frac{\left(a_7-2 a_6\right) V_{us}^*}{\sqrt{6}}$\\\hline
$X^{Bc}_{\eta}\to   \pi^-   J/\psi $ & $ \frac{\left(a_6+a_7\right) V_{ud}^*}{\sqrt{6}}$&
$X^{Bc}_{\eta}\to   K^-   J/\psi $ & $ \frac{(a_6 -2a_7) V_{us}^*}{\sqrt{6}}$\\\hline
\hline
channel & amplitude&channel & amplitude \\\hline
$X^{Bc}_{\pi^0}\to    D^0  D^- $ & $ \frac{\left(a_9-a_{10}\right) V_{ud}^*}{\sqrt{2}}$&
$X^{Bc}_{\pi^0}\to    D^0   D^-_s $ & $ \frac{a_9 V_{us}^*}{\sqrt{2}}$\\\hline
$X^{Bc}_{\pi^+}\to    D^0  \overline D^0 $ & $ \left(a_8+a_{10}\right) V_{ud}^*$&
$X^{Bc}_{\pi^+}\to    D^+  D^- $ & $ \left(a_8+a_9\right) V_{ud}^*$\\\hline
$X^{Bc}_{\pi^+}\to    D^+   D^-_s $ & $ a_9 V_{us}^*$&
$X^{Bc}_{\pi^+}\to    D^+_s   D^-_s $ & $ a_8 V_{ud}^*$\\\hline
$X^{Bc}_{\overline K^0}\to    D^0   D^-_s $ & $ a_{10} V_{ud}^*$&
$X^{Bc}_{K^0}\to    D^0  D^- $ & $ a_{10} V_{us}^*$\\\hline
$X^{Bc}_{K^+}\to    D^0  \overline D^0 $ & $ \left(a_8+a_{10}\right) V_{us}^*$&
$X^{Bc}_{K^+}\to    D^+  D^- $ & $ a_8 V_{us}^*$\\\hline
$X^{Bc}_{K^+}\to    D^+_s  D^- $ & $ a_9 V_{ud}^*$&
$X^{Bc}_{K^+}\to    D^+_s   D^-_s $ & $ \left(a_8+a_9\right) V_{us}^*$\\\hline
$X^{Bc}_{\eta}\to    D^0  D^- $ & $ \frac{\left(a_9+a_{10}\right) V_{ud}^*}{\sqrt{6}}$&
$X^{Bc}_{\eta}\to    D^0   D^-_s $ & $ \frac{(a_9 -2 a_{10}) V_{us}^*}{\sqrt{6}}$\\\hline
\hline
channel & amplitude&channel & amplitude \\\hline
$X^{Bc}_{\pi^-}\to   \pi^-   \pi^-  $ & $ 2 \left(a_{14}+a_{15}\right) V_{ud}^*$&
$X^{Bc}_{\pi^-}\to   \pi^-   K^-  $ & $ \left(a_{14}+a_{15}\right) V_{us}^*$\\\hline
$X^{Bc}_{\pi^0}\to   \pi^0   \pi^-  $ & $ \left(a_{14}+a_{15}\right) V_{ud}^*$&
$X^{Bc}_{\pi^0}\to   \pi^0   K^-  $ & $ \frac{1}{2} \left(a_{12}+2 a_{14}+a_{15}\right) V_{us}^*$\\\hline
$X^{Bc}_{\pi^0}\to   \pi^-   \overline K^0  $ & $ \frac{\left(a_{12}-a_{15}\right) V_{us}^*}{\sqrt{2}}$&
$X^{Bc}_{\pi^0}\to   \pi^-   \eta  $ & $ \frac{\left(a_{12}-a_{13}\right) V_{ud}^*}{\sqrt{3}}$\\\hline
$X^{Bc}_{\pi^0}\to   K^0   K^-  $ & $ \frac{\left(a_{12}-a_{13}\right) V_{ud}^*}{\sqrt{2}}$&
$X^{Bc}_{\pi^0}\to   K^-   \eta  $ & $ \frac{\left(a_{15}-a_{12}\right) V_{us}^*}{2 \sqrt{3}}$\\\hline
$X^{Bc}_{\pi^+}\to   \pi^+   \pi^-  $ & $ \left(2 a_{11}+a_{12}+a_{13}+a_{14}\right) V_{ud}^*$&
$X^{Bc}_{\pi^+}\to   \pi^+   K^-  $ & $ \left(a_{12}+a_{14}\right) V_{us}^*$\\\hline
$X^{Bc}_{\pi^+}\to   \pi^0   \pi^0  $ & $ \left(2 a_{11}+a_{12}+a_{13}-a_{15}\right) V_{ud}^*$&
$X^{Bc}_{\pi^+}\to   \pi^0   \overline K^0  $ & $ \frac{\left(a_{15}-a_{12}\right) V_{us}^*}{\sqrt{2}}$\\\hline
$X^{Bc}_{\pi^+}\to   \pi^0   \eta  $ & $ \frac{\left(a_{13}-a_{12}\right) V_{ud}^*}{\sqrt{3}}$&
$X^{Bc}_{\pi^+}\to   K^+   K^-  $ & $ \left(2 a_{11}+a_{13}\right) V_{ud}^*$\\\hline
$X^{Bc}_{\pi^+}\to   K^0   \overline K^0  $ & $ \left(2 a_{11}+a_{12}\right) V_{ud}^*$&
$X^{Bc}_{\pi^+}\to   \overline K^0   \eta  $ & $ \frac{\left(a_{15}-a_{12}\right) V_{us}^*}{\sqrt{6}}$\\\hline
$X^{Bc}_{\pi^+}\to   \eta   \eta  $ & $ \frac{1}{3} \left(6 a_{11}+a_{12}+a_{13}+a_{15}\right) V_{ud}^*$&
$X^{Bc}_{K^-}\to   \pi^-   K^-  $ & $ \left(a_{14}+a_{15}\right) V_{ud}^*$\\\hline
$X^{Bc}_{K^-}\to   K^-   K^-  $ & $ 2 \left(a_{14}+a_{15}\right) V_{us}^*$&
$X^{Bc}_{\overline K^0}\to   \pi^0   K^-  $ & $ \frac{\left(a_{13}-a_{15}\right) V_{ud}^*}{\sqrt{2}}$\\\hline
$X^{Bc}_{\overline K^0}\to   \pi^-   \overline K^0  $ & $ \left(a_{13}+a_{14}\right) V_{ud}^*$&
$X^{Bc}_{\overline K^0}\to   \overline K^0   K^-  $ & $ \left(a_{14}+a_{15}\right) V_{us}^*$\\\hline
$X^{Bc}_{\overline K^0}\to   K^-   \eta  $ & $ \frac{\left(a_{15}-a_{13}\right) V_{ud}^*}{\sqrt{6}}$&
$X^{Bc}_{K^0}\to   \pi^-   K^0  $ & $ \left(a_{14}+a_{15}\right) V_{ud}^*$\\\hline
$X^{Bc}_{K^0}\to   \pi^-   \eta  $ & $ \sqrt{\frac{2}{3}} \left(a_{13}-a_{15}\right) V_{us}^*$&
$X^{Bc}_{K^0}\to   K^0   K^-  $ & $ \left(a_{13}+a_{14}\right) V_{us}^*$\\\hline
$X^{Bc}_{K^+}\to   \pi^+   \pi^-  $ & $ \left(2 a_{11}+a_{13}\right) V_{us}^*$&
$X^{Bc}_{K^+}\to   \pi^0   \pi^0  $ & $ \left(2 a_{11}+a_{13}\right) V_{us}^*$\\\hline
$X^{Bc}_{K^+}\to   \pi^0   K^0  $ & $ \frac{\left(a_{15}-a_{12}\right) V_{ud}^*}{\sqrt{2}}$&
$X^{Bc}_{K^+}\to   \pi^0   \eta  $ & $ \frac{\left(a_{13}-a_{15}\right) V_{us}^*}{\sqrt{3}}$\\\hline
$X^{Bc}_{K^+}\to   \pi^-   K^+  $ & $ \left(a_{12}+a_{14}\right) V_{ud}^*$&
$X^{Bc}_{K^+}\to   K^+   K^-  $ & $ \left(2 a_{11}+a_{12}+a_{13}+a_{14}\right) V_{us}^*$\\\hline
$X^{Bc}_{K^+}\to   K^0   \overline K^0  $ & $ \left(2 a_{11}+a_{12}\right) V_{us}^*$&
$X^{Bc}_{K^+}\to   K^0   \eta  $ & $ \frac{\left(a_{15}-a_{12}\right) V_{ud}^*}{\sqrt{6}}$\\\hline
$X^{Bc}_{K^+}\to   \eta   \eta  $ & $ \frac{1}{3} \left(6 a_{11}+4 a_{12}+a_{13}-2 a_{15}\right) V_{us}^*$&
$X^{Bc}_{\eta}\to   \pi^0   K^-  $ & $ \frac{\left(a_{12}-2 a_{13}+a_{15}\right) V_{us}^*}{2 \sqrt{3}}$\\\hline
$X^{Bc}_{\eta}\to   \pi^-   \overline K^0  $ & $ \frac{\left(a_{12}-2 a_{13}+a_{15}\right) V_{us}^*}{\sqrt{6}}$&
$X^{Bc}_{\eta}\to   \pi^-   \eta  $ & $ \frac{1}{3} \left(a_{12}+a_{13}+3 a_{14}+a_{15}\right) V_{ud}^*$\\\hline
$X^{Bc}_{\eta}\to   K^0   K^-  $ & $ \frac{\left(a_{12}+a_{13}-2 a_{15}\right) V_{ud}^*}{\sqrt{6}}$&
$X^{Bc}_{\eta}\to   K^-   \eta  $ & $ \frac{1}{6} \left(-a_{12}+2 a_{13}+6 a_{14}+5 a_{15}\right) V_{us}^*$\\\hline
\hline
\end{tabular}
\end{table}

\subsection{$b\to u \bar c d/s$ transition: two-body decays into mesons}

The operator $(\bar ub)(\bar qc)$  can form an anti-symmetric ${\bf  \bar 3}$ and a symmetric ${\bf  6}$ representations.
In the transition $b\to u\bar cs$, the nonzero components of the anti-symmetric tensor $H_{\bar 3}''$ and the symmetric tensor
$H_{ 6}$ are given respectively as
$  (H_{\bar 3}'')^{13} =- (H_{\bar 3}'')^{31} =V_{cs}^*$, $ (H_{\bar 6})^{13}=(H_{\bar 6})^{31} =V_{cs}^*.
$
In the transition $b\to
u\bar cd$, the nonzero components can be obtained by interchanging the subscripts $2\leftrightarrow 3$, and replacing $V_{cs}$ by $V_{cd}$.
Therefor, the effective Hamiltonian at the hadron level for $X_{b\bar c8}$ producing two mesons is constructed as
\begin{eqnarray}
&\mathcal{H}_{eff}&=a_{16} (T_{b\bar c8})^i_j (H_{\bar 3} )^{[jk]}D_i D_k +a_{17}  (T_{b\bar c8})^i_j (H_{6} )^{\{jk\}}D_i D_k.
\end{eqnarray}
Also the Feynman diagrams corresponding with the Hamiltonian above are given in Fig.~\ref{fig:topology-2}(a-d). One then deduce the decay amplitudes for different channels shown in Tab.~\ref{tab:bc8_2antiD}, which leads to the relations for decay widths as
\begin{eqnarray*}
\Gamma(X^{Bc}_{\pi^-}\to D^- D^-_s)= \frac{1}{2}\Gamma(X^{Bc}_{K^-}\to  D^-_s D^-_s)=
2\Gamma(X^{Bc}_{\pi^0}\to \overline D^0 D^-_s),\\
\Gamma(X^{Bc}_{\pi^-}\to D^-D^-)= 2\Gamma(X^{Bc}_{K^-}\to  D^-_sD^-),
\Gamma(X^{Bc}_{\pi^+}\to \overline D^0\overline D^0)= 2\Gamma(X^{Bc}_{\overline K^0}\to \overline D^0 D^-_s),\\
\Gamma(X^{Bc}_{K^+}\to \overline D^0\overline D^0)= 2\Gamma(X^{Bc}_{K^0}\to \overline D^0D^-).
\end{eqnarray*}

\begin{table}
\caption{Tetraquark $X_{b\bar c8}$ decays into two anti-charmed mesons.}\label{tab:bc8_2antiD}\begin{tabular}{|cc|cc|}\hline\hline
channel & amplitude&channel & amplitude \\\hline
$X^{Bc}_{\pi^-}\to   D^-  D^- $ & $ 2 \left(a_{16}+a_{17}\right) V_{cd}^*$&
$X^{Bc}_{\pi^-}\to   D^-   D^-_s $ & $ \left(a_{16}+a_{17}\right) V_{cs}^*$\\\hline
$X^{Bc}_{\pi^0}\to   \overline D^0  D^- $ & $ \sqrt{2} a_{16} V_{cd}^*$&
$X^{Bc}_{\pi^0}\to   \overline D^0   D^-_s $ & $ \frac{\left(a_{16}+a_{17}\right) V_{cs}^*}{\sqrt{2}}$\\\hline
$X^{Bc}_{\pi^+}\to   \overline D^0  \overline D^0 $ & $ 2 \left(a_{17}-a_{16}\right) V_{cd}^*$&
$X^{Bc}_{K^-}\to   D^-   D^-_s $ & $ \left(a_{16}+a_{17}\right) V_{cd}^*$\\\hline
$X^{Bc}_{K^-}\to    D^-_s   D^-_s $ & $ 2 \left(a_{16}+a_{17}\right) V_{cs}^*$&
$X^{Bc}_{\overline K^0}\to   \overline D^0   D^-_s $ & $ \left(a_{17}-a_{16}\right) V_{cd}^*$\\\hline
$X^{Bc}_{K^0}\to   \overline D^0  D^- $ & $ \left(a_{17}-a_{16}\right) V_{cs}^*$&
$X^{Bc}_{K^+}\to   \overline D^0  \overline D^0 $ & $ 2 \left(a_{17}-a_{16}\right) V_{cs}^*$\\\hline
$X^{Bc}_{\eta}\to   \overline D^0  D^- $ & $ \sqrt{\frac{2}{3}} a_{17} V_{cd}^*$&
$X^{Bc}_{\eta}\to   \overline D^0   D^-_s $ & $ \frac{\left(3 a_{16}-a_{17}\right) V_{cs}^*}{\sqrt{6}}$\\\hline
\hline
\end{tabular}
\end{table}

\subsection{Charmless $b\to q_1 \bar q_2 q_3$ transition: two body decays into mesons}

The charmless tree level operator $(\bar q_1 b)(\bar q_2 q_3)$ ($q_i=d,s$) can be decomposed into a triple $H_{\bf 3}$, an antisymmetric sextet $H_{\bf\overline6}$ and a traceless symmetric $H_{\bf{15}}$ in upper indices, while the charmless penguin level operator behave as the triplet $H_{\bf 3}$ .

For the $\Delta S=0 (b\to d)$ decays, the nonzero components of these irreducible tensors are given as
\begin{eqnarray}
 (H_3)^2=1,\;\;\;(H_{\overline6})^{12}_1=-(H_{\overline6})^{21}_1=(H_{\overline6})^{23}_3=-(H_{\overline6})^{32}_3=1,\nonumber\\
 2(H_{15})^{12}_1= 2(H_{15})^{21}_1=-3(H_{15})^{22}_2=
 -6(H_{15})^{23}_3=-6(H_{15})^{32}_3=6.\label{eq:H3615_bd}
\end{eqnarray}
For the $\Delta S=1 (b\to s)$ decays, the nonzero entries in the irreducible tensor $H_{\bf{3}}$, $H_{\bf\overline6}$,
$H_{\bf{15}}$ can be obtained from Eq.~\eqref{eq:H3615_bd}
with the exchange $2\leftrightarrow 3$.
Accordingly, the hadron-level effective Hamiltonian for $X_{b\bar c8}$ decays into mesons is constructed  as
\begin{eqnarray}
&\mathcal{H}_{eff}&=b_4(T_{b\bar c8})^i_j (H_3 )^{k}D_i M^j_k +b_5(T_{b\bar c8})^i_j (H_3 )^{j}D_k M^k_i +b_6 (T_{b\bar c8})^i_j (H_3 )^{k}D_k M^j_i \nonumber\\
&&+b_7(T_{b\bar c8})^i_j (H_{\bar 6} )^{[jk]}_i D_l M^l_k + b_8(T_{b\bar c8})^i_j (H_{\bar 6} )^{[kl]}_i D_k M^j_l + b_9(T_{b\bar c8})^i_j (H_{\bar 6} )^{[jk]}_l D_i M^l_k \nonumber\\
&&+ b_{10}(T_{b\bar c8})^i_j (H_{\bar 6} )^{[jk]}_l D_k M^l_i +b_{11} (T_{b\bar c8})^i_j (H_{15} )^{\{jk\}}_i D_l M^l_k
+ b_{12}(T_{b\bar c8})^i_j (H_{15} )^{\{kl\}}_i D_k M^j_l \nonumber\\
&&+ b_{13}(T_{b\bar c8})^i_j (H_{15} )^{\{jk\}}_l D_i M^l_k + b_{14}(T_{b\bar c8})^i_j (H_{15} )^{\{jk\}}_l D_k M^l_i.
\end{eqnarray}
In the two-body decays of the transition, the decay amplitudes are given in Tab.~\ref{tab:bc8_Dbar_Md} for the transition $b\to d$ and Tab.~\ref{tab:bc8_Dbar_Ms} for the transition $b\to s$. We obtain no direct relation of these decay widths.
\begin{table}
\caption{Tetraquark $X_{b\bar c8}$ decays into anti-charmed and light mesons induced by the charmless $b\to d$ transition.}\label{tab:bc8_Dbar_Md}\begin{tabular}{|c|c|c|c|c|c|c|c}\hline\hline
channel & amplitude \\\hline
$X^{Bc}_{\pi^-}\to   D^-  \pi^-  $ & $ b_4+b_6+b_9+b_{10}-2 b_{12}+3 b_{13}+3 b_{14}$\\\hline
$X^{Bc}_{\pi^0}\to   \overline D^0  \pi^-  $ & $ \frac{1}{\sqrt{2}} \left( b_4- b_5+ b_7+ b_8+ b_9+ b_{10}+5  b_{11}+3  b_{12}+3 b_{13}-3  b_{14}\right)$\\\hline
$X^{Bc}_{\pi^0}\to   D^-  \pi^0  $ & $ \frac{1}{2} \left(b_4+b_5+2 b_6-b_7-b_8+b_9+b_{10}-5 b_{11}+b_{12}-5 b_{13}+b_{14}\right)$\\\hline
$X^{Bc}_{\pi^0}\to   D^-  \eta  $ & $ \frac{-b_4-b_5+b_7-b_8+3 b_9+b_{10}+5 b_{11}+5 b_{12}-3 b_{13}+5 b_{14}}{2 \sqrt{3}}$\\\hline
$X^{Bc}_{\pi^0}\to    D^-_s  K^0  $ & $ \frac{-b_5+b_7-b_{10}+5 b_{11}+b_{14}}{\sqrt{2}}$\\\hline
$X^{Bc}_{\pi^+}\to   \overline D^0  \pi^0  $ & $ -\frac{b_4-b_5+b_7+b_8+b_9+b_{10}-3 b_{11}+3 b_{12}-5 b_{13}-3 b_{14}}{\sqrt{2}}$\\\hline
$X^{Bc}_{\pi^+}\to   \overline D^0  \eta  $ & $ \frac{b_4+b_5-b_7+b_8-3 b_9-b_{10}+3 b_{11}+3 b_{12}+3 b_{13}+3 b_{14}}{\sqrt{6}}$\\\hline
$X^{Bc}_{\pi^+}\to   D^-  \pi^+  $ & $ b_5+b_6-b_7-b_8+3 b_{11}+3 b_{12}-2 b_{14}$\\\hline
$X^{Bc}_{\pi^+}\to    D^-_s  K^+  $ & $ b_5-b_7+b_{10}+3 b_{11}-b_{14}$\\\hline
$X^{Bc}_{K^-}\to   D^-  K^-  $ & $ b_6+b_8+b_{10}-b_{12}+3 b_{14}$\\\hline
$X^{Bc}_{K^-}\to    D^-_s  \pi^-  $ & $ b_4-b_8+b_9-b_{12}+3 b_{13}$\\\hline
$X^{Bc}_{\overline K^0}\to   \overline D^0  K^-  $ & $ b_5+b_7-b_{10}-b_{11}+3 b_{14}$\\\hline
$X^{Bc}_{\overline K^0}\to   D^-  \overline K^0  $ & $ b_5+b_6+b_7+b_8-b_{11}-b_{12}-2 b_{14}$\\\hline
$X^{Bc}_{\overline K^0}\to    D^-_s  \pi^0  $ & $ \frac{-b_4+b_8-b_9+b_{12}+5 b_{13}}{\sqrt{2}}$\\\hline
$X^{Bc}_{\overline K^0}\to    D^-_s  \eta  $ & $ \frac{1}{\sqrt{6}} \left( b_4-2  b_5-2  b_7- b_8-3  b_9-2  b_{10}+2 b_{11}- b_{12}+3 b_{13}+2 b_{14}\right)$\\\hline
$X^{Bc}_{K^0}\to   D^-  K^0  $ & $ b_4+b_6-b_9-b_{10}-2 b_{12}-b_{13}-b_{14}$\\\hline
$X^{Bc}_{K^+}\to   \overline D^0  K^0  $ & $ b_4+b_8-b_9+3 b_{12}-b_{13}$\\\hline
$X^{Bc}_{K^+}\to   D^-  K^+  $ & $ b_6-b_8-b_{10}+3 b_{12}-b_{14}$\\\hline
$X^{Bc}_{\eta}\to   \overline D^0  \pi^-  $ & $ \frac{b_4+b_5+3 b_7+b_8+b_9-b_{10}+3 b_{11}+3 b_{12}+3 b_{13}+3 b_{14}}{\sqrt{6}}$\\\hline
$X^{Bc}_{\eta}\to   D^-  \pi^0  $ & $ -\frac{b_4+b_5+3 b_7+b_8+b_9-b_{10}+3 b_{11}-5 b_{12}-5 b_{13}-5 b_{14}}{2 \sqrt{3}}$\\\hline
$X^{Bc}_{\eta}\to   D^-  \eta  $ & $ \frac{1}{6} \left(b_4+b_5+3 \left(2 b_6+b_7+b_8-b_9-b_{10}+b_{11}-b_{12}+b_{13}-b_{14}\right)\right)$\\\hline
$X^{Bc}_{\eta}\to    D^-_s  K^0  $ & $ \frac{-2 b_4+b_5+3 b_7+2 b_8+2 b_9+b_{10}+3 b_{11}+2 b_{12}+2 b_{13}-b_{14}}{\sqrt{6}}$\\\hline
\hline
\end{tabular}
\end{table}
\begin{table}
\caption{Tetraquark $X_{b\bar c8}$ decays into anti-charmed and light mesons induced by the charmless $b\to s$ transition .}\label{tab:bc8_Dbar_Ms}\begin{tabular}{|cc|cc|c|c|c|c|c|c}\hline\hline
channel & amplitude&channel & amplitude \\\hline
$X^{Bc}_{\pi^-}\to   D^-  K^-  $ & $ b_4-b_8+b_9-b_{12}+3 b_{13}$&
$X^{Bc}_{\pi^-}\to    D^-_s  \pi^-  $ & $ b_6+b_8+b_{10}-b_{12}+3 b_{14}$\\\hline
$X^{Bc}_{\pi^0}\to   \overline D^0  K^-  $ & $ \frac{b_4+2 b_7+b_8+b_9+4 b_{11}+3 b_{12}+3 b_{13}}{\sqrt{2}}$&
$X^{Bc}_{\pi^0}\to   D^-  \overline K^0  $ & $ \frac{ \left(- b_4+2  b_7+b_8+b_9+4  b_{11}+ b_{12}+ b_{13}\right)}{\sqrt{2}}$\\\hline
$X^{Bc}_{\pi^0}\to    D^-_s  \pi^0  $ & $ b_6+b_{12}+b_{14}$&
$X^{Bc}_{\pi^0}\to    D^-_s  \eta  $ & $ \frac{-2 b_7-b_8+b_{10}-4 b_{11}+2 b_{12}+2 b_{14}}{\sqrt{3}}$\\\hline
$X^{Bc}_{\pi^+}\to   \overline D^0  \overline K^0  $ & $ b_4+b_8-b_9+3 b_{12}-b_{13}$&
$X^{Bc}_{\pi^+}\to    D^-_s  \pi^+  $ & $ b_6-b_8-b_{10}+3 b_{12}-b_{14}$\\\hline
$X^{Bc}_{K^-}\to    D^-_s  K^-  $ & $ b_4+b_6+b_9+b_{10}-2 b_{12}+3 b_{13}+3 b_{14}$&
$X^{Bc}_{\overline K^0}\to    D^-_s  \overline K^0  $ & $ b_4+b_6-b_9-b_{10}-2 b_{12}-b_{13}-b_{14}$\\\hline
$X^{Bc}_{K^0}\to   \overline D^0  \pi^-  $ & $ b_5+b_7-b_{10}-b_{11}+3 b_{14}$&
$X^{Bc}_{K^0}\to   D^-  \pi^0  $ & $ -\frac{b_5+b_7+2 b_9+b_{10}-b_{11}-4 b_{13}-b_{14}}{\sqrt{2}}$\\\hline
$X^{Bc}_{K^0}\to   D^-  \eta  $ & $ \frac{-2 b_4+b_5+b_7+2 b_8+b_{10}-b_{11}+2 b_{12}+6 b_{13}-b_{14}}{\sqrt{6}}$&
$X^{Bc}_{K^0}\to    D^-_s  K^0  $ & $ b_5+b_6+b_7+b_8-b_{11}-b_{12}-2 b_{14}$\\\hline
$X^{Bc}_{K^+}\to   \overline D^0  \pi^0  $ & $ \frac{ \left(b_5-b_7-2  b_9- b_{10}+3  b_{11}+4  b_{13}+3  b_{14}\right)}{\sqrt{2}}$&
$X^{Bc}_{K^+}\to   \overline D^0  \eta  $ & $ -\frac{2 b_4-b_5+b_7+2 b_8+b_{10}-3 b_{11}+6 b_{12}-6 b_{13}-3 b_{14}}{\sqrt{6}}$\\\hline
$X^{Bc}_{K^+}\to   D^-  \pi^+  $ & $ b_5-b_7+b_{10}+3 b_{11}-b_{14}$&
$X^{Bc}_{K^+}\to    D^-_s  K^+  $ & $ b_5+b_6-b_7-b_8+3 b_{11}+3 b_{12}-2 b_{14}$\\\hline
$X^{Bc}_{\eta}\to   \overline D^0  K^-  $ & $ \frac{b_4-2 b_5+b_8+b_9+2 b_{10}+6 b_{11}+3 b_{12}+3 b_{13}-6 b_{14}}{\sqrt{6}}$&
$X^{Bc}_{\eta}\to   D^-  \overline K^0  $ & $ \frac{b_4-2 b_5-b_8-b_9-2 b_{10}+6 b_{11}-b_{12}-b_{13}+2 b_{14}}{\sqrt{6}}$\\\hline
$X^{Bc}_{\eta}\to    D^-_s  \pi^0  $ & $ \frac{-b_8+2 b_9+b_{10}+2 b_{12}-4 b_{13}+2 b_{14}}{\sqrt{3}}$&
$X^{Bc}_{\eta}\to    D^-_s  \eta  $ & $ \frac{2 b_4}{3}+\frac{2 b_5}{3}+b_6-2 b_{11}-b_{12}-2 b_{13}-b_{14}$\\\hline
\hline
\end{tabular}
\end{table}

\subsection{ $\bar c\to \bar q_1 q_2 \bar q_3$  transition: two body decays into mesons}
Under the flavor SU(3) symmetry, the operator $\bar c q_1  \bar q_2 q_3$  transforms
 as ${\bf  \bar3}\otimes {\bf 3}\otimes {\bf
\bar3}={\bf  \bar3}\oplus {\bf  \bar3}\oplus {\bf6}\oplus {\bf \overline {15}}$.
Following the classifications mentioned before, the Cabibbo allowed transition to be $\bar c\to \bar s  d \bar u$, and the nonzero tensor components are given as
\begin{eqnarray}
(H_{ 6})_{31}^2=-(H_{6})_{13}^2=1,\;\;\;
 (H_{\overline {15}})_{31}^2= (H_{\overline {15}})_{13}^2=1.\label{eq:H3615_c_allowed}
\end{eqnarray}
In  the  singly Cabibbo suppressed transition $\bar c\to \bar u d\bar d$ and $\bar c\to \bar u  s\bar s$ , the combination of tensor components are given as
\begin{eqnarray}
(H_{6})_{31}^3 =-(H_{6})_{13}^3 =(H_{ 6})_{12}^2 =-(H_{ 6})_{21}^2 =\sin(\theta_C),\nonumber\\
 (H_{\overline {15}})_{31}^3= (H_{\overline {15}})_{13}^3=-(H_{\overline {15}})_{12}^2=-(H_{\overline {15}})_{21}^2= \sin(\theta_C).\label{eq:H3615_cc_singly_suppressed}
\end{eqnarray}
while for the doubly Cabibbo suppressed transition  $\bar c\to \bar d  s \bar u$, we have
\begin{eqnarray}
(H_{ 6})_{21}^3=-(H_{ 6})_{12}^3=\sin^2\theta_C,\;\;
 (H_{\overline {15}})_{21}^3= (H_{\overline {15}})_{12}^3=\sin^2\theta_C. \label{eq:H3615_c_doubly_suprressed}
\end{eqnarray}
Therefor, it is convenient to construct the hadron-level effective Hamiltonian for $X_{b\bar c8}$ decays into mesons  as
\begin{eqnarray}
&\mathcal{H}_{eff}&=f_3(T_{b\bar c8})^i_j (H_6)^j_{[ik]} \overline B^l M^k_l + f_4(T_{b\bar c8})^i_j (H_6)^l_{[ik]} \overline B^j M^k_l +f_5 (T_{b\bar c8})^i_j (H_6)^l_{[ik]} \overline B^k M^j_l \nonumber\\
&&+ f_6(T_{b\bar c8})^i_j (H_6)^j_{[kl]} \overline B^k M^l_i +f_7(T_{b\bar c8})^i_j (H_{\overline {15}})^j_{\{ik\}} \overline B^l M^k_l +f_8 (T_{b\bar c8})^i_j (H_{\overline {15}})^l_{\{ik\}} \overline B^j M^k_l \nonumber\\
&&+ f_9(T_{b\bar c8})^i_j (H_{\overline {15}})^l_{\{ik\}} \overline B^k M^j_l +f_{10} (T_{b\bar c8})^i_j (H_{\overline {15}})^j_{\{kl\}} \overline B^k M^l_i.
\end{eqnarray}
As usual, the corresponding Feynman diagrams are given in Fig.~\ref{fig:topology-4}. Expanding the Hamiltonian above to obtain decay amplitudes, and listed in Tab.~\ref{tab:bc8_B_M}. One deduce the relations between different decay widths given as
\begin{eqnarray*}
    \Gamma(X^{Bc}_{\pi^-}\to B^-\pi^- )= { }\Gamma(X^{Bc}_{K^-}\to B^-K^- ),
    \Gamma(X^{Bc}_{\pi^0}\to B^-\pi^0 )= { }\Gamma(X^{Bc}_{\eta}\to B^-\eta ),\\
    \Gamma(X^{Bc}_{\pi^+}\to B^-\pi^+ )= { }\Gamma(X^{Bc}_{K^+}\to B^-K^+ ),
    \Gamma(X^{Bc}_{\pi^+}\to \overline B^0_s\pi^0 )= { }\Gamma(X^{Bc}_{\pi^0}\to \overline B^0_s\pi^- ),\\
     \Gamma(X^{Bc}_{K^0}\to B^-K^0 )= { }\Gamma(X^{Bc}_{\overline K^0}\to B^-\overline K^0 ),
     \Gamma(X^{Bc}_{K^0}\to \overline B^0_s\pi^- )= { }\Gamma(X^{Bc}_{\overline K^0}\to \overline B^0K^- ),\\
     \Gamma(X^{Bc}_{K^+}\to \overline B^0K^0 )= { }\Gamma(X^{Bc}_{\pi^+}\to \overline B^0_s\overline K^0 ).
\end{eqnarray*}

\begin{table}
\caption{Tetraquark $X_{b\bar c8}$ decays into a B meson and light meson.}\label{tab:bc8_B_M}\begin{tabular}{|cc|cc|c|c|c|c|c|c}\hline\hline
channel & amplitude&channel & amplitude \\\hline
$X^{Bc}_{\pi^0}\to   B^-  K^0  $ & $ \frac{-f_4+f_6+f_8-f_{10}}{\sqrt{2}}$&
$X^{Bc}_{\pi^0}\to   \overline B^0_s  \pi^-  $ & $ -\frac{f_5+f_6-f_9+f_{10}}{\sqrt{2}}$\\\hline
$X^{Bc}_{\pi^+}\to   B^-  K^+  $ & $ -f_3-f_6+f_7+f_{10}$&
$X^{Bc}_{\pi^+}\to   \overline B^0  K^0  $ & $ -f_3-f_4+f_7+f_8$\\\hline
$X^{Bc}_{\pi^+}\to   \overline B^0_s  \pi^0  $ & $ \frac{f_5+f_6-f_9+f_{10}}{\sqrt{2}}$&
$X^{Bc}_{\pi^+}\to   \overline B^0_s  \eta  $ & $ \frac{ \left(2  f_3- f_5+ f_6-2  f_7+ f_9+ f_{10}\right)}{\sqrt{6}}$\\\hline
$X^{Bc}_{K^-}\to   B^-  \pi^-  $ & $ f_4+f_5+f_8+f_9$&
$X^{Bc}_{\overline K^0}\to   B^-  \pi^0  $ & $ \frac{f_3-f_5+f_7-f_9}{\sqrt{2}}$\\\hline
$X^{Bc}_{\overline K^0}\to   B^-  \eta  $ & $ \frac{f_3+f_5+2 f_6+f_7+f_9-2 f_{10}}{\sqrt{6}}$&
$X^{Bc}_{\overline K^0}\to   \overline B^0  \pi^-  $ & $ f_3+f_4+f_7+f_8$\\\hline
$X^{Bc}_{\overline K^0}\to   \overline B^0_s  K^-  $ & $ f_3+f_6+f_7+f_{10}$&
$X^{Bc}_{K^+}\to   \overline B^0_s  K^0  $ & $ -f_4-f_5+f_8+f_9$\\\hline
$X^{Bc}_{\eta}\to   B^-  K^0  $ & $ -\frac{f_4+2 f_5+f_6-f_8+2 f_9-f_{10}}{\sqrt{6}}$&
$X^{Bc}_{\eta}\to   \overline B^0_s  \pi^-  $ & $ \frac{-2 f_4-f_5+f_6-2 f_8+f_9+f_{10}}{\sqrt{6}}$\\\hline
\hline
$X^{Bc}_{\pi^-}\to   B^-  \pi^-  $ & $ \left(f_4+f_5+f_8+f_9\right) (-\text{sC})$&
$X^{Bc}_{\pi^0}\to   B^-  \pi^0  $ & $ \frac{ \left(f_3-f_4-f_5+f_6+f_7+f_8-f_9-f_{10}\right) \text{sC}}{2}$\\\hline
$X^{Bc}_{\pi^0}\to   B^-  \eta  $ & $ \frac{\left(f_3+3 f_4+f_5-f_6+f_7-3 f_8+f_9+f_{10}\right) \text{sC}}{2 \sqrt{3}}$&
$X^{Bc}_{\pi^0}\to   \overline B^0  \pi^-  $ & $ \frac{ \left(f_3+ f_4+ f_5+ f_6+ f_7+ f_8- f_9+ f_{10}\right) \text{sC}}{\sqrt{2}}$\\\hline
$X^{Bc}_{\pi^0}\to   \overline B^0_s  K^-  $ & $ \frac{\left(f_3-f_5+f_7+f_9\right) \text{sC}}{\sqrt{2}}$&
$X^{Bc}_{\pi^+}\to   B^-  \pi^+  $ & $ \left(f_3+f_6-f_7-f_{10}\right) \text{sC}$\\\hline
$X^{Bc}_{\pi^+}\to   \overline B^0  \pi^0  $ & $ -\frac{\left(f_3+f_4+f_5+f_6-f_7-f_8-f_9+f_{10}\right) \text{sC}}{\sqrt{2}}$&
$X^{Bc}_{\pi^+}\to   \overline B^0  \eta  $ & $ \frac{\left(f_3+3 f_4+f_5-f_6-f_7-3 f_8-f_9-f_{10}\right) \text{sC}}{\sqrt{6}}$\\\hline
$X^{Bc}_{\pi^+}\to   \overline B^0_s  \overline K^0  $ & $ \left(f_3-f_5-f_7+f_9\right) \text{sC}$&
$X^{Bc}_{K^-}\to   B^-  K^-  $ & $ \left(f_4+f_5+f_8+f_9\right) \text{sC}$\\\hline
$X^{Bc}_{\overline K^0}\to   B^-  \overline K^0  $ & $ \left(f_5+f_6+f_9-f_{10}\right) \text{sC}$&
$X^{Bc}_{\overline K^0}\to   \overline B^0  K^-  $ & $ \left(f_4-f_6+f_8-f_{10}\right) \text{sC}$\\\hline
$X^{Bc}_{K^0}\to   B^-  K^0  $ & $ \left(f_5+f_6+f_9-f_{10}\right) (-\text{sC})$&
$X^{Bc}_{K^0}\to   \overline B^0_s  \pi^-  $ & $ \left(f_4-f_6+f_8-f_{10}\right) (-\text{sC})$\\\hline
$X^{Bc}_{K^+}\to   B^-  K^+  $ & $ \left(f_3+f_6-f_7-f_{10}\right) (-\text{sC})$&
$X^{Bc}_{K^+}\to   \overline B^0  K^0  $ & $ \left(f_3-f_5-f_7+f_9\right) (-\text{sC})$\\\hline
$X^{Bc}_{K^+}\to   \overline B^0_s  \pi^0  $ & $ \frac{\left(-f_4+f_6+f_8+f_{10}\right) \text{sC}}{\sqrt{2}}$&
$X^{Bc}_{K^+}\to   \overline B^0_s  \eta  $ & $ \frac{ \left(2  f_3+3  f_4+2  f_5+ f_6-2  f_7-3  f_8-2 f_9+ f_{10}\right) \text{sC}}{\sqrt{6}}$\\\hline
$X^{Bc}_{\eta}\to   B^-  \pi^0  $ & $ -\frac{\left(3 f_3+f_4-f_5+f_6+3 f_7-f_8-f_9-f_{10}\right) \text{sC}}{2 \sqrt{3}}$&
$X^{Bc}_{\eta}\to   B^-  \eta  $ & $ -\frac{\left(f_3-f_4-f_5+f_6+f_7+f_8-f_9-f_{10}\right) \text{sC}}{2} $\\\hline
$X^{Bc}_{\eta}\to   \overline B^0  \pi^-  $ & $ -\frac{\left(3 f_3+f_4-f_5+f_6+3 f_7+f_8+f_9+f_{10}\right) \text{sC}}{\sqrt{6}}$&
$X^{Bc}_{\eta}\to   \overline B^0_s  K^-  $ & $ -\frac{ \left(3  f_3+2  f_4+ f_5+2  f_6+3  f_7+2 f_8-f_9+2 f_{10}\right) \text{sC}}{\sqrt{6}}$\\\hline
\hline
$X^{Bc}_{\pi^-}\to   B^-  K^-  $ & $ \left(f_4+f_5+f_8+f_9\right) \text{sC}^2$&
$X^{Bc}_{\pi^0}\to   B^-  \overline K^0  $ & $ -\frac{\left(f_4+f_5-f_8+f_9\right) \text{sC}^2}{\sqrt{2}}$\\\hline
$X^{Bc}_{\pi^0}\to   \overline B^0  K^-  $ & $ -\frac{\left(f_4+f_5+f_8-f_9\right) \text{sC}^2}{\sqrt{2}}$&
$X^{Bc}_{\pi^+}\to   \overline B^0  \overline K^0  $ & $ \left(f_4+f_5-f_8-f_9\right) \left(-\text{sC}^2\right)$\\\hline
$X^{Bc}_{K^0}\to   B^-  \pi^0  $ & $ \frac{\left(f_3+f_6+f_7-f_{10}\right) \text{sC}^2}{\sqrt{2}}$&
$X^{Bc}_{K^0}\to   B^-  \eta  $ & $ \frac{\left(f_3-2 f_5-f_6+f_7-2 f_9+f_{10}\right) \text{sC}^2}{\sqrt{6}}$\\\hline
$X^{Bc}_{K^0}\to   \overline B^0  \pi^-  $ & $ \left(f_3+f_6+f_7+f_{10}\right) \text{sC}^2$&
$X^{Bc}_{K^0}\to   \overline B^0_s  K^-  $ & $ \left(f_3+f_4+f_7+f_8\right) \text{sC}^2$\\\hline
$X^{Bc}_{K^+}\to   B^-  \pi^+  $ & $ \left(f_3+f_6-f_7-f_{10}\right) \left(-\text{sC}^2\right)$&
$X^{Bc}_{K^+}\to   \overline B^0  \pi^0  $ & $ \frac{1}{\sqrt{2}} \left( f_3+ f_6- f_7+ f_{10}\right) \text{sC}^2$\\\hline
$X^{Bc}_{K^+}\to   \overline B^0  \eta  $ & $ -\frac{\left(f_3-2 f_5-f_6-f_7+2 f_9-f_{10}\right) \text{sC}^2}{\sqrt{6}}$&
$X^{Bc}_{K^+}\to   \overline B^0_s  \overline K^0  $ & $ \left(f_3+f_4-f_7-f_8\right) \left(-\text{sC}^2\right)$\\\hline
$X^{Bc}_{\eta}\to   B^-  \overline K^0  $ & $ -\frac{\left(f_4-f_5-2 f_6-f_8-f_9+2 f_{10}\right) \text{sC}^2}{\sqrt{6}}$&
$X^{Bc}_{\eta}\to   \overline B^0  K^-  $ & $ \frac{\left(f_4-f_5-2 f_6+f_8+f_9-2 f_{10}\right) \text{sC}^2}{\sqrt{6}}$\\\hline
\hline
\end{tabular}
\end{table}

\section{Golden Decay Channels}
\label{sec:golden_channels}

In this  section, we will discuss the golden channels to reconstruct the  $X_{b\bar c8}$ and give an estimate of the decay branching fractions.
In our analysis given in the previous sections,  the  final meson can  be replaced by its corresponding counterpart  with the same quark constituent but  different quantum numbers. For instance, one may replace  $\overline K^0$ by $\overline K^{*0}$.

Golden decay  channels must satisfy the following criteria.
\begin{itemize}
\item Branching fractions:  For charm quark decays, one should use the Cabibbo allowed decay modes, while for bottom quark, the quark level transition  $b\to c\bar ud$ or $b\to c\bar cs$ gives the largest branching fractions.

\item Detection efficiency: At hadron colliders like LHC,  charged particles have better chances to be detected than neutral states. So we will remove the channels with   $\pi^0$, $\eta$, $\phi$, $\rho^{\pm}(\to \pi^{\pm}\pi^0$), $K^{*\pm}(\to K^{\pm}\pi^0$) and $\omega$, but keep the modes with $\pi^\pm, K^0(\to \pi^+\pi^-), \rho^0(\to \pi^+\pi^-)$.

\end{itemize}


\begin{table}
 \caption{Cabibbo allowed $X_{b\bar c8}$ $\bar c$-quark decays.  }\label{tab:Xb6_golden_meson_c}\begin{tabular}{|c  c   c   c c|}\hline\hline
$X^{Bc}_{\pi^0}\to   B^-  K^0  $&
$X^{Bc}_{\pi^0}\to   \overline B^0_s  \pi^-  $& & &\\
\hline
$X^{Bc}_{\pi^+}\to   B^-  K^+  $&
$X^{Bc}_{\pi^+}\to   \overline B^0  K^0  $& & &\\
\hline
$X^{Bc}_{K^-}\to B^- l^-\bar\nu $&$X^{Bc}_{K^-}\to   B^-  \pi^-  $&& &\\
\hline
$X^{Bc}_{\overline K^0}\to \overline B^0 l^-\bar\nu $&$X^{Bc}_{\overline K^0}\to   \overline B^0  \pi^-  $&
$X^{Bc}_{\overline K^0}\to   \overline B^0_s  K^-  $& &\\

\hline
$X^{Bc}_{K^+}\to   \overline B^0_s  K^0  $&&&&\\
\hline
$X^{Bc}_{\eta}\to \overline B^0_s l^-\bar\nu $&$X^{Bc}_{\eta}\to   B^-  K^0  $&
$X^{Bc}_{\eta}\to   \overline B^0_s  \pi^-  $&&\\
\hline\hline
\end{tabular}
\end{table}

\begin{table}
 \caption{Golden channels for  $X_{b\bar c8}$ $b$-quark decays.  }\label{tab:Xb6_golden_meson_b}\begin{tabular}{|c  c   c  c c|}\hline\hline
$X^{Bc}_{\pi^-}\to   \pi^-    D^-_s $&
$X^{Bc}_{\pi^-}\to   K^-   D^- $&
$X^{Bc}_{\pi^-}\to   \pi^-   \pi^-  $&
$X^{Bc}_{\pi^-}\to   D^-   D^-_s $& \\

\hline
$X^{Bc}_{\pi^0}\to   \pi^-   J/\psi $&
$X^{Bc}_{\pi^0}\to   \overline K^0   D^- $&
$X^{Bc}_{\pi^0}\to   K^-   \overline D^0 $&
$X^{Bc}_{\pi^0}\to    D^0  D^- $&
$X^{Bc}_{\pi^0}\to   K^0   K^-  $\\
$X^{Bc}_{\pi^0}\to   \overline D^0   D^-_s $&
& & & \\
\hline
$X^{Bc}_{\pi^+}\to   \pi^+    D^-_s $&
$X^{Bc}_{\pi^+}\to   \overline K^0   \overline D^0 $&
$X^{Bc}_{\pi^+}\to    D^0  \overline D^0 $&
$X^{Bc}_{\pi^+}\to    D^+  D^- $& \\
$X^{Bc}_{\pi^+}\to    D^+_s   D^-_s $&
$X^{Bc}_{\pi^+}\to   \pi^+   \pi^-  $&
$X^{Bc}_{\pi^+}\to   K^+   K^-  $&
$X^{Bc}_{\pi^+}\to   K^0   \overline K^0  $&\\

\hline
$X^{Bc}_{K^-}\to   K^-    D^-_s $&
$X^{Bc}_{K^-}\to   \pi^-   K^-  $&
$X^{Bc}_{K^-}\to    D^-_s   D^-_s $&&\\

\hline

$X^{Bc}_{\overline K^0}\to   K^-   J/\psi $&
$X^{Bc}_{\overline K^0}\to   \overline K^0    D^-_s $&
$X^{Bc}_{\overline K^0}\to    D^0   D^-_s $&
$X^{Bc}_{\overline K^0}\to   \pi^-   \overline K^0  $&\\

\hline
$X^{Bc}_{K^0}\to   D^-  J/\psi $&
$X^{Bc}_{K^0}\to   \pi^-   \overline D^0 $&
$X^{Bc}_{K^0}\to   K^0    D^-_s $&
$X^{Bc}_{K^0}\to   \pi^-   K^0  $&
$X^{Bc}_{K^0}\to   \overline D^0  D^- $\\

\hline

$X^{Bc}_{K^+}\to   \overline D^0  J/\psi $&
$X^{Bc}_{K^+}\to   K^0   J/\psi $&
$X^{Bc}_{K^+}\to   \pi^+   D^- $&
$X^{Bc}_{K^+}\to   K^+    D^-_s $&
$X^{Bc}_{K^+}\to    D^+_s  D^- $\\
$X^{Bc}_{K^+}\to   \pi^-   K^+  $&
$X^{Bc}_{K^+}\to   \overline D^0  \overline D^0 $&
 & &\\
\hline
$X^{Bc}_{\eta}\to    D^-_s  J/\psi $&
$X^{Bc}_{\eta}\to   \pi^-   J/\psi $&
$X^{Bc}_{\eta}\to   \overline K^0   D^- $&
$X^{Bc}_{\eta}\to   K^-   \overline D^0 $&
$X^{Bc}_{\eta}\to    D^0  D^- $\\
$X^{Bc}_{\eta}\to   K^0   K^-  $&
$X^{Bc}_{\eta}\to   \overline D^0   D^-_s $&&
&\\
\hline\hline
\end{tabular}
\end{table}

The  two-body  decay modes that can be used to reconstruct the $X_{b\bar c8}$  are collected in Tab.~\ref{tab:Xb6_golden_meson_c} and Tab.~\ref{tab:Xb6_golden_meson_b}.

Some comments are appropriate to our SU(3) analysis. The typical branching fraction for charm quark decay in Tab.~\ref{tab:Xb6_golden_meson_c} is at a few percent level. On the experiment side, to construct the bottom meson, another factor, at the order  $10^{-3}$ or even smaller,  due to the weak decay of bottom meson is needed. So the  branching fraction for the decay chains to reconstruct the $X_{b\bar c8}$ might reach  the order $10^{-5}$, or smaller.

If the $b$ quark decay first in $X_{b\bar c8}$, the typical branching fraction is at the order $10^{-3}$. The final states which include the $J/\psi$ or $D$ meson such as $X^{Bc}_{\pi^-}\to   K^-   D^- $ would introduce  a factor  $10^{-3}$ to reconstruct. So the branching fraction of these channels may also reach  the order $10^{-5}$.

The channels with two light mesons such as $X^{Bc}_{\pi^-}\to   \pi^-   \pi^-  $ require the annihilation of the two heavy quarks. But since the CKM matrix element $V_{cb}$, these channels might have  sizable decay branching fractions.


\section{Conclusions}
\label{sec:conclusions}

Tetraquarks with the quark content  $[b q_i] [\bar {c}\bar {q_j}]$ are of great research interest but have  not  been discovered yet.
In this paper, we have systematically studied the weak decays of the doubly-heavy tetraquarks $X_{b\bar c8}$  under SU(3) flavor symmetry, which include  the semileptonic and nonleptonic $b$ and $\bar c$ quark decays. The $\bar c$ quark decays are dominant, of which the typical branching fraction is at a few percents level.

Using the building blocks in SU(3), we construct the effective  Hamiltonian  at the hadron level for their weak decays. The nonperturbative effects are parametrized into a few quantities  ($a_{i},b_{j},...$). Therefore, one can easily derive the decay amplitudes, based on which  relations between different channels can be obtained. Finally, we give a list of  the golden channels which is  useful to search  for the $X_{b\bar c8}$ state tetraquarks in future experiments.

\section*{Acknowledgments}

We thank Prof. Wei Wang for useful discussions.
 This work is
supported in part by   National Natural Science Foundation of
China under Grant No.~11575110, 11655002, 11675091, 11735010, and 11835015, by Natural Science Foundation of Shanghai under Grant  No.~15DZ2272100,  by Key
Laboratory for Particle Physics, Astrophysics and Cosmology,
Ministry of Education.

\end{document}